\documentclass[manuscript]{aastex}

\usepackage{lscape}
\usepackage{longtable}
\usepackage{ctable}
\usepackage{multirow}
\usepackage{amssymb,amsmath}

\newcommand{\bb}{\begin{equation}}
\newcommand{\ee}{\end{equation}}
\newcommand{\s}[1]{{_{\rm{#1}}}}

\shorttitle{Modeling Pulsar Wind Nebulae}
\shortauthors{Vorster and Tibolla}

\begin{document}

\title{Time-dependent Modeling of Pulsar Wind Nebulae}

   \author{M.J. Vorster}
   \affil{Centre for Space Research, North-West University, Potchefstroom Campus, 2520 Potchefstroom, South Africa\\
       \email{12792322@puk.ac.za} }
       
   \author{O. Tibolla}
   \affil{Institut f\"ur Theoretische Physik und Astrophysik, Universit\"at W\"urzburg, Campus Hubland Nord, Emil-Fischer-Str. 31, D-97074 W\"urzburg, Germany \\
       \email{omar.tibolla@gmail.com} }

   \author{S.E.S. Ferreira}
   \affil{Centre for Space Research, Potchefstroom Campus, North-West University, 2520 Potchefstroom, South Africa}

  \author{S. Kaufmann}
  \affil{Landessternwarte, Universit\"at Heidelberg, K\"onigstuhl 12, D-69117 Heidelberg, Germany}

\begin{abstract}
A spatially independent model that calculates the time evolution of the electron spectrum in a spherically expanding pulsar wind nebula (PWN) is presented,  allowing one to make broadband predictions for the PWN's non-thermal radiation.   The source spectrum of electrons injected at the termination shock of the PWN is chosen to be a broken power law.  In contrast to previous PWN models of a similar nature, the source spectrum has a discontinuity in intensity at the transition between the low and high-energy components.  To test the model, it is applied to the young PWN G21.5--0.9, where it is found that a discontinuous source spectrum can model the emission at all wavelengths better than a continuous one.  The model is also applied to the unidentified sources HESS J1427--608 and HESS J1507--622.  Parameters are derived for these two candidate nebulae that are consistent with the values predicted for other PWNe.  For HESS J1427--608 a present-day magnetic field of $B\s{age}=0.4\,\mu\text{G}$ is derived.  As a result of the small present-day magnetic field, this source has a low synchrotron luminosity, while remaining bright at GeV/TeV energies.  It is therefore possible to interpret HESS J1427--608 within the ancient PWN scenario.  For the second candidate PWN HESS J1507--622, a present-day magnetic field of $B\s{age}=1.7\,\mu\text{G}$ is derived.  Furthermore, for this candidate PWN a scenario is favoured in the present paper in which HESS J1507--622 has been compressed by the reverse shock of the supernova remnant.  
\end{abstract}

\keywords{ISM: individual objects (G21.5--0.9, HESS J1427--608, HESS J1507--622) -- ISM: supernova remnants -- radiation mechanisms: non-thermal}

\section{Introduction}

Pulsars produce highly relativistic winds that consist of electrons and positrons, and possibly a hadronic component \citep[e.g.,][]{Cheng1980, Bednarek2003}.  Due to the nature of the wind, the ideal magnetohydrodynamic limit is satisfied, leading to the pulsar's magnetic field being frozen into the out-flowing wind \citep[e.g.,][]{Kirk2009}.  When the ram pressure of the wind is equal to the confining pressure of the ambient medium, a termination shock is formed \citep{Rees1974} where the charged particles are re-accelerated \citep[e.g.,][]{Reynolds1984}.   

Downstream of the termination shock the electrons (and positrons) interact with the frozen-in magnetic field, leading to synchrotron radiation that is observed from radio to X-ray wavelengths.  Additionally, the electrons can also inverse Compton (IC) scatter ambient photons to high-energy (HE) and very-high-energy (VHE) gamma-ray wavelengths.  These ambient photons can have a number of origins, including the cosmic microwave background radiation (CMBR), infra-red (IR) radiation from dust, starlight, and even the radiated synchrotron photons.  The non-thermal emission leads to a luminous nebula, commonly known as a pulsar wind nebula (PWN). 

A growing number of PWNe are observed that are located inside shell-type supernova remnants (SNRs), with this type of system commonly known as composite remnants.  Notable examples include the PWN G21.5--0.9 \citep[e.g.,][]{Bocchino2005}, and the Vela PWN \citep[e.g.,][]{Bock1998}.  The presence of a shell component plays an important role in determining the morphological evolution of a PWN, which in turn influences the non-thermal emission.  The evolution of the nebula can broadly be divided into three phases: 
\begin{itemize}
\item
In the initial phase the pulsar injects energy at a constant rate into the nebula, resulting in a PWN that expands supersonically into the slow-moving stellar material that was ejected during the preceding supernova explosion \citep[e.g.,][]{Vanderswaluw2001, Gaensler2006}.  Theoretical models predict that the expansion of the outer boundary of the PWN can be described by $R\s{pwn} \propto t^{1.1-1.2}$,  while the average magnetic field decreases as $B\propto t^{-1.3}$ \citep[e.g.,][]{Reynolds1984}.
\item
The shock wave of the SNR consists of both a forward and reverse shock.  The reverse shock initially expands outwards along with the forward shock, but when the pressure in the remnant become sufficiently small, the reverse shock will start to propagate towards the centre of the shell \citep{McKee1974}.  The PWN will enter the next evolutionary phase when the reverse shock reaches $R\s{pwn}$.  The reverse shock compresses the PWN \citep[e.g.,][]{Vanderswaluw2001, Bucciantini2003}, leading to a larger magnetic field, and consequently larger synchrotron losses and a brighter radio/X-ray nebula \citep[e.g.,][]{Reynolds1984}.  
\item
After the compression of $R\s{pwn}$, the PWN enters a second expansion phase.  In contrast to the initial expansion phase, $R\s{pwn}(t)$ does not evolve smoothly, but has an oscillatory nature \citep[e.g.,][]{Vanderswaluw2001, Bucciantini2003}, i.e., the PWN goes through various contractions and expansions.  As the ejecta surrounding the PWN has been heated by the reverse shock, the expansion of $R\s{pwn}$ in this phase is subsonic.  If the energy output of the pulsar has significantly declined at the onset of the last phase, then the conservation of magnetic flux implies that an expanding nebula will lead to a decrease in the average magnetic field.  As a result, the synchrotron component of the non-thermal emission will grow fainter with time.     
\end{itemize}     

It is well-known that the X-ray synchrotron emission observed from PWNe is produced by a young population of electrons, as these particles have a relatively short lifetime \citep[e.g.,][]{Shklovsky1957}.  The evolution of the X-ray emission is therefore correlated with the evolution of the magnetic field, which in turn is dependent on the morphological evolution of the nebula.  By contrast, the electrons producing VHE gamma-ray emission have a significantly longer lifetime, implying that the TeV emission observed from PWNe is produced by particles that have accumulated over the lifetime of the pulsar \citep[e.g.,][]{Dejager2009}.  This is strikingly illustrated by the energy-dependent morphology of the $\sim 21\,\text{kyr}$ old nebula HESS J1825-137, where VHE gamma-ray observations reveal a PWN that is significantly larger than the associated X-ray nebula \citep{H_Aharonian2006}.  For a PWN with an average magnetic field of $B=5\,\mu\text{G}$, the lifetime of an electron emitting $1\,\text{keV}$ X-rays is $\sim 3\,\text{kyr}$, whereas the corresponding lifetime of an electron producing $1\,\text{TeV}$ gamma-rays is $\sim 19\,\text{kyr}$ \citep[e.g.,][]{Dejager2009}.

Based on the information presented in the previous paragraphs, \citet{Dejager2008a} proposed that the average magnetic field in an aged PWN could evolve below the $B\sim 3\,\mu\text{G}$ value of the interstellar medium (ISM), resulting in these sources being undetectable at synchrotron frequencies.  However, due to the longer lifetimes of the VHE gamma-ray producing electrons, these ancient PWNe may still be visible at TeV energies.  As PWNe count among the more common TeV sources, the ancient PWN scenario could offer an explanation for a number of unidentified TeV sources that lack a synchrotron counterpart \citep{H_Aharonian2008} .  

Two unidentified sources that were proposed by \cite{Tibolla2011} as ancient PWNe candidates are HESS J1427--608 \citep{H_Aharonian2008} and HESS J1507--622 \citep{H_Aharonian2011}, and in this paper we extend the initial time-dependent modeling of \cite{Tibolla2011} for these sources.  Before introducing the model, the two-component nature of the PWN electron spectrum is briefly reviewed in Section \ref{sec:two_com}, as this has some implications for the model.  In Section \ref{sec:model}, the spatially independent model used to calculate the time evolution of the electron spectrum is presented.   In order to test the model, it is first applied to the young PWN G21.5--0.9, with the modeling results discussed at the beginning of Section \ref{sec:results}.  This section also focuses on modeling HESS J1427--608 and HESS J1507--622 within a PWN framework.  The aim is to not only investigate the ancient PWN hypothesis, but also to determine whether a clear argument can be made for identifying HESS J1427--608 and HESS J1507--622 as PWNe.  The final section of the paper deals with the discussion and main conclusions drawn from the modeling.

\section{The two-component electron spectrum}\label{sec:two_com}

Observations of PWNe indicate that the energy spectrum of the electrons (and positrons) responsible for the non-thermal emission can be separated into two distinct components \citep[e.g.,][]{Weiler1978, Gaensler2006, Dejager2009}: (1) a low-energy component producing the radio synchrotron and GeV IC emission, and (2), a high-energy component producing the X-ray synchrotron and TeV IC emission.  It further follows from observations that these two components can both be described by a power-law $N_e\propto E^{-\alpha}$, with $\alpha_R \sim 1.0-1.3$ for the low-energy component \citep[e.g.,][]{Weiler1978}, and $\alpha_X \sim 2$ for the high-energy component.  The stated value of $\alpha_X$ is specifically in the vicinity of the shock, as synchrotron losses and diffusion will lead to an evolution of $\alpha_X$ as the particles propagate away from the shock \citep[][]{Bocchino2005, Mangano2005, Schock2010}.  A similar evolution is not expected for the low-energy component, as diffusion and synchrotron losses are markedly more effective at higher energies.  For a discussion on the spatial evolution of the particle spectra in PWNe, \citet{Vorster2013} can be consulted.  Although it has never been measured directly, particle evolution models predict that the transition between the two components should occur at an energy of $E\lesssim 0.3\,\text{TeV}$ \citep{Zhang2008, Fang2010, Tanaka2011}.    

Motivated by the above-mentioned considerations, particle evolution models often use a broken power-law to describe the spectrum of electrons injected into the PWN at the termination shock \citep[e.g.,][]{Venter2006, Zhang2008, Tanaka2011}.  While these models typically assume that the two components connect smoothly, i.e., have the same intensity at the transition, \citet{Dejager2008c} concluded from their modeling of Vela X that the two components do not connect smoothly, but that the low-energy component should cut off steeply in order to connect to the high-energy component.  In this paper it will be demonstrated that this discontinuity is not limited to the particle spectrum of the aged Vela PWN, but also seems to be present in the spectrum of the young nebula G21.5--0.9.

The question naturally arises as to how these two electron populations are formed.  As demonstrated by \citet{Axford1977}, \citet{Krymskii1977}, \citet{Bell1978}, and \citet{Blandford1978}, diffusive shock acceleration leads to a power law energy spectrum $N_e \propto E^{-\alpha}$, with $\alpha\geq 2$.  It is therefore possible to associate the origin of the high-energy component with this process.  To explain the low-energy component is more difficult, as $\alpha_R=1$, and one would not naturally associate the origin of this component with diffusive shock acceleration.  \citet{Summerlin2012} recently showed that it is nevertheless possible for relativistic magnetohydrodynamic shocks to produce this hard spectrum if particles are subjected to shock drift acceleration.  

An alternative explanation for the origin of the low-energy component has also been proposed by \citet{Spitkovsky2008}.  Results from particle-in-cell simulations show that the acceleration of particles at the termination shock leads to a relativistic Maxwellian spectrum with a non-thermal power-law tail.  This result is also indirectly supported by the modeling of \citet{Fang2010}, and \citet{Grondin2011}, where these authors were able to reproduce the non-thermal emission from four PWNe using the spectrum predicted by \citet{Spitkovsky2008}.  The advantage of this spectrum is that it provides a natural explanation for the discrepancy in intensity between the two components.  A Maxwellian spectrum will therefore also be used for the modeling of G21.5--0.9.  

Extending the simulations of \citet{Spitkovsky2008}, \citet{Sironi2011} found that magnetic reconnection occurring in the striped pulsar wind \citep[e.g.,][]{Coroniti1990} can accelerate particles at the termination shock, leading to a deviation from a Maxwellian spectrum.  From their simulations it follows that this modified low-energy component can be described by a power-law with $\alpha_R \sim 1.5$.  These results may be supported by the observations of \citet{Dodson2003}.  Focusing on the very inner regions of the Vela PWN, these authors found radio lobes in the equatorial plane of the nebula.  As magnetic reconnection will also occur around the equatorial plane, it is possible that these lobes are formed by the accelerated particles that have been injected into the nebula.  

In summary, it is clear that two distinct electron populations should be present in PWNe.  However, the exact nature of these components is still unknown.

\section{The Model}\label{sec:model}

The temporal evolution of the electron spectrum in a PWN can be calculated using the equation \citep[e.g.,][]{Tanaka2010}
\begin{equation}\label{eq:dn_dt}
\frac{\partial N_e(E_e,t)}{\partial t} = Q(E_e,t) + \frac{\partial}{\partial E}\left[\dot{E}(E_e,t)N_e(E_e,t)\right], 
\end{equation}
where $E_e$ represents the electron energy and $N_e(E_e,t)$ the number of electrons per energy interval.  The number of electrons injected into the PWN at the termination shock, per time and energy interval, is given by $Q(E_e,t)$, while the second term on the right-hand side of Equation (\ref{eq:dn_dt}) describes continuous energy losses (or gains) suffered by the particles, with $\dot{E}(E_e,t)$ the total energy loss rate.   

Emulating \citet{Venter2006}, a broken power-law spectrum is used to model the emission from the sources studied in this paper,
\bb\label{eq:power-law}
Q(E_e,t) = \begin{cases}
Q_{\rm{R}}\left(E_{\rm{b}}/E_e\right), & \text{if } E_{\min} \le E_e \le E_{\rm{b}}\\
Q_{\rm{X}}\left(E_{\rm{b}}/E_e\right)^2, & \text{if } E_{\rm{b}} < E_e \le E_{\max} 
\end{cases},
\ee
where $Q_{\rm{R}}$ and $Q_{\rm{X}}$ are normalisation constants, $E_{\min}$ and $E_{\max}$ the minimum and maximum electron energy, respectively, and $E_{\rm{b}}$ the energy where the spectrum transitions between the two components.  Note that the indices chosen for the components of the broken power-law follow from the discussion presented in Section \ref{sec:two_com}.  Keeping in mind that \citet{Dejager2008c} derived a discontinuous spectrum for Vela X, it is not an a priori requirement that the two components should have the same intensity at $E\s{b}$.

The normalisation constants are determined by the prescription that the total energy in a given component should be some fraction $\eta_i$ ($i=\mbox{ R,X}$) of the pulsar's spin-down luminosity $L(t)$ \citep[e.g.,][]{Venter2006}
\bb\label{eq:Q_calc}
\int Q_i\left(E_{\rm{b}}/E_e\right)^{p_i} E_e dE_e = \eta_i L(t).
\ee
Assuming that the pulsar is a pure dipole radiator with a braking index of $3$, the time-dependence of the luminosity is given by
\bb\label{eq:L_t}
L(t) = \frac{L_0}{\left(1+t/\tau\right)^2}.
\ee
In the expression above $L_0$ represents the initial luminosity and $\tau$ the characteristic spin-down time scale of the pulsar.

The total energy loss rate in the model, $\dot{E}$ in Equation (\ref{eq:dn_dt}), includes both synchrotron radiation and IC scattering, as well as adiabatic cooling/heating.  The energy loss rate as a result of synchrotron radiation and IC scattering is given by \citep[e.g.,][]{Longair2011}
\bb\label{eq:E_dot_syn}
\dot{E}\s{n-t}(E_e,t) = \frac{4}{3}\frac{\sigma\s{T}}{\left(m_e c\right)^2 c}E_e^2  U_B\left(1+\frac{U\s{IC}}{U_B}\right),
\ee
where $\sigma\s{T}$ is the Thomson cross-section, $U_B=B^2/8\pi$ the energy density of the magnetic field, and $U\s{IC}$ the energy density of the target photon field.  Although Equation (\ref{eq:E_dot_syn}) describes IC scattering in the Thomson regime, \citet{Moderski2005} have shown that this expression is still valid if $U_{\rm{IC}}/U_B \lesssim 3$, and Klein-Nishina effects can be neglected.  For the CMBR with an energy density of $U\s{IC}\sim 0.3\,\text{eV}\,\text{cm}^{-3}$, this condition is satisfied for an average magnetic field of $B > 2\,\mu\text{G}$.  

Anticipating the modeling results, it was however found that this condition is violated for both HESS J1427--608 and HESS J1507--622, and Klein-Nishina effects are therefore taken into account by multiplying Equation (\ref{eq:E_dot_syn}) with a correction factor $F\s{KN}$.  \citet{Moderski2005} have further shown that when the target photon field is described by a black-body spectrum, the modification factor can be approximated by
\bb\label{eq:KN_correction}
F\s{KN} \sim \frac{1}{\left(1+4\gamma\epsilon\right)},
\ee    
where $\gamma=E_e/m_ec^2$ is the Lorentz factor of the particle, and $\epsilon = 2.8k\s{B} T/m_ec^2$.  In the last expression $k\s{B}$ represents Boltzmann's constant and $T$ the temperature of the black-body spectrum.  The approximation given in Equation (\ref{eq:KN_correction}) is appropriate as the photon fields used for the modeling of the unidentified TeV sources are described by a black-body spectrum.          

For adiabatic cooling (heating), the energy loss rate is given by  
\bb\label{eq:E_dot_ad}
\dot{E}\s{ad}(E_e,t) = \frac{1}{3}\left(\nabla \cdot \mathbf{v}\right)E_e,
\ee
where $\mathbf{v}$ is the convection velocity downstream of the termination shock.  If the system has a spherical symmetry, then Equation (\ref{eq:E_dot_ad}) can be simplified to
\bb\label{eq:ad_r}
\dot{E}\s{ad}(E_e,t) = \frac{1}{3r^2}\frac{\partial}{\partial r}\left[r^2 v(r)\right]E_e.
\ee  
From the above expression it follows that particles suffer the largest amount of adiabatic losses in the inner part of the system.  Furthermore, if $v\propto 1/r^2$, then the particles will suffer no adiabatic losses.  Therefore, to correctly include adiabatic losses requires $v(r)$ to be known.  \citet{Tanaka2010} included adiabatic losses in their spatially independent PWN using the approximation
\bb\label{eq:E_dot_ad2}
\dot{E}\s{ad}(E_e,t) = \frac{v\s{pwn}(t)}{R\s{pwn}(t)}E_e, 
\ee
where $v\s{pwn}(t)$ and $R\s{pwn}(t)$ are respectively the expansion velocity and radius of the PWN.  This approximation is also used for the present model.      

As discussed in the Introduction, the PWN goes through three evolutionary phases, with the expansion/compression of $R\s{pwn}(t)$ approximated using the power-laws  
\bb\label{eq:R_t}
R_{\rm{pwn}}(t) = \begin{cases}
R_0(t/t_0)^{r_1} & \text{if } t<t_{\rm{rs}}\\
R_0(t_{\rm{rs}}/t_0)^{r_1}(t/t_{\rm{rs}})^{r_2} & \text{if } t_{\rm{rs}}\le t<t_{\rm{se}}\\
R_0(t_{\rm{rs}}/t_0)^{r_1}(t\s{se}/t_{\rm{rs}})^{r_2}(t/t_{\rm{se}})^{r_3} & \text{if } t\ge t_{\rm{se}}
\end{cases}.
\ee
Here $t\s{rs}$ represents the time needed for the reverse shock of the SNR to reach the PWN, and $t\s{se}$ the time when the PWN enters the second expansion phase.  For the initial condition, $R_0=0.01\,\text{pc}$ when $t_0=10\,\text{yr}$ \citep{Gelfand2009}.  The values $r_1$, $r_2$, and $r_3$ are not linearly independent, as the size of the PWN predicted by the model must be equal to the observed size.  Note that the distance to the source $d$ influences the values of $r_1$, $r_2$ and $r_3$, as a larger value of $d$ implies a larger source, and hence a faster expansion.  As a point of reference, \citet{Reynolds1984} calculated that $R_{\rm{pwn}} \propto t^{1.2}$ when $t < \tau$.  

Apart from $R\s{pwn}$, the adiabatic loss rate described by Equation (\ref{eq:E_dot_ad2}) is also a function of $v\s{pwn}(t)=dR_{\rm{pwn}}(t)/dt$.  \citet{Gelfand2009} calculated that the expansion velocity increases from $v\s{pwn}(t)\sim 1300\,\text{km}\,\text{s}^{-1}$ at $t=0.01\,\text{kyr}$ to $v\s{pwn}(t)\sim 2300\,\text{km}\,\text{s}^{-1}$ at $t=5\,\text{kyr}$.  However, these values were calculated for a specific scenario, and are only provided as a point of reference.  

The reverse shock time scale $t_{\rm{rs}}$ is given by \cite[e.g.][]{Reynolds1984, Ferreira2008}
\bb\label{eq:t_rev}
t_{\rm{rs}} = 4 \left(\frac{M_{\rm{ej}}}{3M_{\odot}}\right)^{3/4} \left(\frac{E_{\rm{ej}}}{10^{51}\,\text{erg}}\right)^{-45/100} \biggl(\frac{n_{\rm{ism}}}{ 1\,\text{cm}^{-3}}\biggr)^{-1/3}\,\text{kyr},
\ee
where $M_{\rm{ej}}$ and $E_{\rm{ej}}$ are respectively the mass and kinetic energy of the supernova ejecta, and $n_{\rm{ism}}$ the particle number density of ISM.  Using the fiducial value of $E_{\rm{ej}}=10^{51}\,\text{erg}$, along with the values of $M_{\rm{ej}}=5 M_{\odot}$ and $n_{\rm{ism}}= 1\,\text{cm}^{-3}$, leads to the estimate $t\s{rs} \approx 6\,\text{kyr}$.  Using a smaller value for the ISM density, $n_{\rm{ism}}= 0.1\,\text{cm}^{-3}$, increases the reverse shock time scale to $t\s{rs} \approx 11\,\text{kyr}$.   

The evolution of the average magnetic field in the nebula $B\s{pwn}(t)$ is calculated using the conservation of magnetic flux \citep[e.g.,][]{Tanaka2010}
\bb\label{eq:B_evolve}
\int_0^{t} \eta_B L(t)dt = V\s{pwn}(t)\frac{B^2_{\rm{pwn}}(t)}{8\pi},
\ee 
where $\eta_B$ is the fraction of the pulsar's spin-down luminosity converted into magnetic energy, and $V\s{pwn}(t)$ the volume of the PWN.  Using $R_{\rm{pwn}} \propto t^{1.2}$, along with the fact that $L$ is effectively time-independent when $t < \tau$, leads to the time evolution of the magnetic field $B \propto t^{-1.3}$, identical to the time-dependence derived by \cite{Reynolds1984}.  

An important parameter in the study of PWNe is the ratio of electromagnetic to particle energy in the nebula $\sigma$.  In terms of the present model, this ratio is defined as 
\bb
\sigma = \frac{\eta_B}{\eta\s{R}+\eta\s{X}},
\ee
and is subjected to the constraint $\sigma \lesssim 1$ \citep[e.g.,][]{Dejager2009}.  An additional constraint follows from $\eta_B+\eta\s{R}+\eta\s{X} \lesssim 1$.  This sum is not set strictly equal to unity to allow for the fact that a fraction $\eta\s{rad}$ of the pulsar's spin-down luminosity is radiated away in the form of pulsed emission, i.e., $\eta_B+\eta\s{R}+\eta\s{X}+\eta\s{rad} = 1$.  The value of $\eta\s{rad}$ is difficult to determine, but the results from the first \emph{Fermi}-LAT pulsar catalogue \citep{Abdo2010} suggest that $\eta\s{rad}\sim 1\%-10\%$ is reasonable.  For the modeling $\eta\s{rad} \lesssim 1\%$ is used, similar to the value derived for the Crab pulsar \citep{Abdo2010}.  Note that this small value used for $\eta\s{rad}$ in the model effectively implies that $\sigma \simeq \eta_B$.  

Apart from energy losses, the model also takes into account that particles can escape from the PWN as a result of diffusion.  The escape time scale $\tau\s{esc}$ is given by   \citep{Parker1965}
\bb\label{eq:tau_esc}
\tau\s{esc}(t) = \frac{R^2_{\rm{pwn}}(t)}{6\kappa(t)},
\ee
where $\kappa(t)$ is the diffusion coefficient.  Diffusion in a PWN results from particles interacting with irregularities in the magnetic field, and it may be argued that $\kappa(t) \propto 1/B\s{pwn}(t)$ \citep[e.g.,][]{Lerche1981}.  Furthermore, the diffusion coefficient is chosen to scale linearly with energy, i.e., $\kappa\equiv \kappa (E_e/1\mbox{ TeV})$.  This functional form of $\kappa$ is similar to the form derived for Bohm diffusion
\bb\label{eq:kappa_bohm}
\kappa\s{Bohm} = \frac{cE_e}{3qB},
\ee  
where $q$ is the electric charge of the particle.

The temporal evolution of the electron spectrum is not obtained by solving Equation (\ref{eq:dn_dt}) directly, but rather in the following fashion: the amount of particles with energy $E_e$ injected into the PWN over the time interval $dt$ is given by $Q(E_e,t)dt$, where $Q(E_e,t)$ is specified using Equation (\ref{eq:power-law}).  The injected particles are then added to the current value of $N_e(E,t)$ to obtain the total number of particles in the nebula $N_e(E_e,t+dt)=N_e(E_e,t)+Q(E_e,t)dt$.  In the time interval $dt$ the particles also suffer energy losses as a result of adiabatic cooling and non-thermal radiation.  The new energy of the particles is given by $E_e'=E_e-dE_e$, where $dE_e$ is the sum of the loss rates given in Equations (\ref{eq:E_dot_syn}) and (\ref{eq:E_dot_ad2}).  The particles $N_e(E_e,t+dt)$ will therefore evolve to a new position in energy space $N'_e(E_e',t+dt)$.  During the interval $dt$ a fraction of the particles will also have escaped from the nebula.  Assuming that the particles are distributed uniformly throughout the nebula, this fraction is given by $\xi\s{esc}=dt/\tau\s{esc}$, with the escape time scale calculated using Equation (\ref{eq:tau_esc}).  A value $\xi\s{esc} \ge 1$ indicates that all particles have effectively escaped from the system.  Note that if $\xi\s{esc} > 1$, then the value $\xi\s{esc} = 1$ is used in the model.  The number of particles remaining in the nebula after a time $t+dt$ is thus given by $N'_e(E_e',t+dt)(1-\xi\s{esc})$.  This approach is essentially similar to the those used by, e.g., \citet{Gelfand2009}, \citet{Schock2010}, and \citet{Vanetten2011}, and is adequate provided that $dt$ is chosen sufficiently small.

In order to determine if the model predicts the correct evolution of the electron spectrum, it was tested using four well-known criteria: (1) adiabatic losses only lead to a reduction in the intensity of the spectrum, but does not lead to any spectral changes \citep[e.g.,][]{Vorster2013}; (2) if the system is in a steady-state, synchrotron losses lead to a spectrum that is steeper by one power of $E_e$ when compared to the source spectrum \citep[e.g.,][]{Pacholczyk1970}; (3) adiabatic losses primarily affect the low- energy electrons, while synchrotron losses are more important for the high-energy electrons \citep[e.g.,][]{Vorster2013}; and (4) in the absence of synchrotron losses, particles escaping from a system will result in a softer spectrum, with this softening directly related to the energy dependence of the diffusion coefficient \citep[e.g.,][]{Lerche1981}.  In the present model $\kappa \propto E_e$, and the spectrum should again be softer by one power of $E_e$ (compared to the source spectrum).  It was found that the model predicts all of the above-mentioned behaviour.

In reality, the evolution of a PWN could be considerably more complex than described by the model.  One might, for example, think of a nebula that expands in a very inhomogeneous ISM.  Simulations by, e.g., \citet{Blondin2001} and \citet{Vorster2013b}, show that in such a scenario the reverse shock of the SNR will be asymmetric, leading to a cigar or bullet-shaped PWN.  One might also argue, with merit, that a more realistic model should include a spatial dependence, like the model presented by \citet{Vorster2013}.  However, in the case where only spatially integrated observations are available, it is not clear how useful a spatially-dependent model would be.  In this regard, the present model should be viewed as a first-order approximation.  Furthermore, any time dependence in the conversion efficiencies $\eta\s{R}$, $\eta\s{X}$, and $\eta\s{B}$, as well as the values of the energy spectrum $E\s{min}$, $E\s{max}$, and $E\s{b}$, is not taken into account.   It is unknown, to the authors at least, if any theoretical calculations exist that predict the time-dependence of the above-mentioned parameters.   

The non-thermal emission is calculated using the appropriate expressions given in \cite{Blumenthal1970}.  This implies that Klein-Nishina effects are taken into account when calculating the IC spectrum.

\section{Results}\label{sec:results}

\subsection{A Test-case PWN}

\begin{figure}[!ht]
\begin{center}
\includegraphics[width=0.85\textwidth]{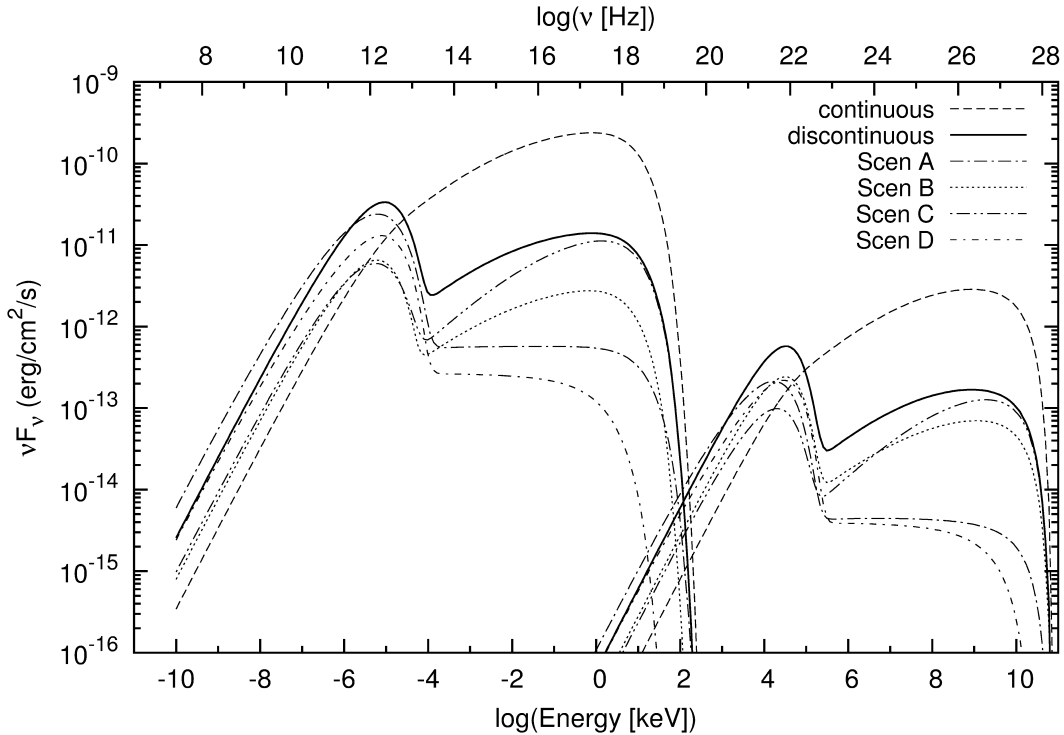}
\caption{Model PWN showing the influence of the various parameters on the evolution of the non-thermal radiation spectra.  The discontinuous spectrum is used as a reference scenario (see text for parameter values used), while Scenarios A--D are identical to the basis scenario, with the exception of one varied parameter.  The parameters that were varied are listed in Table \ref{tab:param_vary}.}
\label{fig:test}
\end{center}
\end{figure}

It is instructive to first apply the model to a general PWN in order to illustrate the effect that the various free parameters have on the evolution of the non-thermal radiation spectra.  For the test case, a PWN placed at a distance of $1\,\text{kpc}$ is allowed to expand for $t\s{age}=1\,\text{kyr}$, with the nebula having a present-day size of $R\s{pwn}=1\,\text{pc}$, and an expansion velocity of $v\s{age}=1000\,\text{km}\,\text{s}^{-1}$.  The conversion efficiency of spin-down luminosity to magnetic energy is chosen to be $\eta_B=0.03$, leading to a present-day magnetic field of $B\s{age}=50\,\mu\text{G}$.  For this specific PWN, both adiabatic and escape losses are neglected.      

The first important distinction that can be made is the effect of a continuous/discontinuous particle spectrum on the broadband emission.  Note that both these source spectra are described by Equation (\ref{eq:power-law}), but with the following difference: for the continuous spectrum the two components of the broken power-law have the same intensity at $E\s{b}$.  In the case of the discontinuous spectrum, the high-energy component has a lower intensity, compared to the low-energy component, at $E\s{b}$.  The continuous spectrum requires only a single conversion efficiency, and consequently a single conversion efficiency $\eta\equiv\eta_R+\eta_X$.  For both the continuous and discontinuous spectra the values $\eta_R=0.92$ and $\eta_X=0.05$ are chosen, implying that the same fraction of the pulsar's spin-down luminosity is converted to particle energy for the two cases.

The radiation spectra obtained with the two types of source spectra are shown in Figure \ref{fig:test}, from which it can be seen that the continuity/discontinuity in the particle spectra is preserved in the non-thermal emission.  Another salient feature visible in Figure \ref{fig:test} is that the different source spectra predict fluxes for the non-thermal emission that differ by more than an order of magnitude.  This discrepancy is easy to explain if it is kept in mind that in both scenarios the same amount of spin-down energy is converted to particle energy.  To transform a discontinuous source spectrum into continuous spectrum requires that a fraction of the energy stored in the low-energy component must be transferred to the high-energy component, thereby reducing the intensity of the former component, while increasing the intensity of the latter component.    As the low-energy component is described by $N_e\propto E_e^{-1}$, and the high-energy component by $N_e\propto E_e^{-2}$, a small change in the intensity (or equivalently energy) of the former component will lead to a larger change in the intensity of the latter component.  This can also be seen from Figure \ref{fig:test} where the discrepancy between the X-ray spectra is larger than the discrepancy between the radio spectra.  With these differences taken into account, it is clear that the two source spectra will lead to the derivation of different parameters for the same PWN.  This will be discussed in a more qualitative fashion when the modelling results of G21.5--0.9 are presented in Section \ref{sec:results_G21.5}.     

Having demonstrated the effect of the source spectrum on the non-thermal emission, it is also necessary to demonstrate the effect that different values for the various parameters have on the evolution of the non-thermal spectra.  For this purpose a number of alternative scenarios are chosen, with these alternative scenarios having only one varied parameter compared to the discontinuous source spectrum scenario.  The parameters varied are listed in Table \ref{tab:param_vary}, with the quantity in brackets indicating the value of the reference scenario.

\begin{table}[t]
\textbf{\caption[]{\label{tab:param_vary}
{\textnormal{Parameter values used for the scenarios depicted in Figure \ref{fig:test}.  The value in brackets indicates the value used for the discontinuous source spectrum scenario.}}}}
\begin{center}
	\begin{tabular}{ll}
	\hline\midrule
 	Scenario & Difference to reference scenario \\
  \midrule
Scen A & PWN has an age of $t\s{age}=10\,\text{kyr}$ ($t\s{age}=1\,\text{kyr}$) \\ 
Scen B & spin-down time scale of the pulsar is $\tau=0.3\,\text{kyr}$ ($\tau=1\,\text{kyr}$)\\ 
Scen C & adiabatic losses are included (adiabatic losses are neglected)\\ 
Scen D & escape losses are included (escape losses are neglected) \\ 
  \midrule
\end{tabular}
\end{center}
\end{table}

The first quantity varied is the age of the system, corresponding to Scen A in Figure \ref{fig:test}.  As the PWN ages, the intensity of the X-ray and TeV spectra decrease as a result of synchrotron and IC losses, while the spectral index becomes steeper by one power of energy.  By contrast, the low-energy component remains largely unaffected by non-thermal losses.  As mentioned at the end of Section \ref{sec:model}, this is the theoretically expected effect of non-thermal losses, thus indicating that the model works correctly.  

In the next scenario, Scen B, the spin-down time scale is reduced to $\tau=0.3\,\text{kyr}$.  Figure \ref{fig:test} shows that the influence of this parameter is to reduce the non-thermal flux at all wavelengths.  At times $t < \tau$, the spin-down luminosity of the pulsar is approximately constant, but decreases rapidly after $t > \tau$.  As the number of particles injected into the nebula is determined by the time integral over $L$, which in turn is dependent on the value of $\tau$, a smaller number of particles have been injected into the nebula in Scen B (compared to the reference scenario), leading to the reduced flux.  

The next parameter investigated is the effect of adiabatic losses on the radiation spectra, Scen C.  It can be seen from Figure \ref{fig:test} that the low-energy component of the particle spectrum is primarily affected by these losses, leading to a decrease in the intensity of the radio/GeV spectra.  For the high-energy component of the source spectrum, synchrotron losses are more important, and the effect of adiabatic losses becomes negligible.  One noteworthy point is that adiabatic losses do not affect the spectral index.  

Lastly, Scen D shows the effect of escape losses on the radiation spectra.  As $\kappa$ scales with energy, the high-energy component of the particle spectrum is primarily affected, leading to a decrease in the intensity of the X-ray/TeV spectra.  One important feature of escape losses in a spatially independent model is that this process leads to a spectral evolution that is very similar to that of synchrotron losses, and one might argue that it would not be possible to distinguish between the two loss processes.  Therefore, to find a model prediction that is compatible with the data, the escape losses are initially fixed using the Bohm diffusion coefficient (\ref{eq:kappa_bohm}) while all other parameters are varied.  Only after a reasonable agreement has been found is the diffusion coefficient varied to improve the model prediction.  The same is also true for the parameters $E\s{min}$, $E\s{max}$, and $E\s{b}$.

\begin{table}[!ht]
\textbf{\caption[]{\label{tab:parameters}
{\textnormal{Values derived with the model for the various free parameters.  Values marked with an * represent parameters that were kept fixed, or parameters that follow from the derived model parameters.}}}}
\begin{center}
	\begin{tabular}{lcccc}
	\hline\midrule
 	Parameter & Symbol & G21.5--0.9 & J1427--608 & J1507--622 \\
  \midrule
  Initial spin-down luminosity ($10^{38}\,\text{erg}\,\text{s}^{-1}$) & $L_0$ & $0.54*$ & $5.5$ & $1.2$ \\
  Spin-down time scale (kyr) & $\tau$ & $3^*$ & $3$ & $0.5$ \\
  Age of nebula (kyr) & $t\s{age}$ & $0.87^*$ & $10$ & $24$ \\
  Present-day magnetic field ($\mu\text{G}$) & $B\s{age}$ & $230$ & $0.4$ & $1.7$  \\
  Radio conversion efficiency & $\eta\s{R}$ & $0.68$ & $0.81$ & $0.8$ \\
  X-ray conversion efficiency & $\eta\s{X}$ & $0.13$ & $0.18$ & $0.17$ \\
  Ratio: particle conversion efficiencies & $\eta\s{R}/\eta\s{X}$ & $5.4^*$ & $4.5^*$ & $4.7^*$ \\
  Ratio: magnetic to particle energy ($10^{-3}$) & $\sigma$ & $180$ & $0.01$ & $30$  \\
  Minimum electron energy ($10^{-3}\,\text{TeV}$) & $E\s{min}$ & $0.3$ & $100$ & $1$ \\
  Maximum electron energy ($10^2\,\text{TeV}$) & $E\s{max}$ & $2.7$ & $3$ & $2$ \\
  Break energy (TeV) & $E\s{b}$ & $0.1$ & $0.18$ & $0.5$ \\  
  Diffusion coefficient ($10^{25}  E\s{TeV}\,\text{cm}^2\,\text{s}^{-1}$) & $\kappa$ & $2.2$ & $7$ & $15$\\
  Ratio: diffusion coefficients & $\kappa/\kappa\s{Bohm}$ & $390^*$ & $2.3^*$ & $20^*$\\
  Distance to source (kpc) & $d$ & $4.8^*$ & $11^*$ & $6^*$ \\
  \midrule
\end{tabular}
\end{center}
\end{table}


\subsection{G21.5--0.9}\label{sec:results_G21.5}

\begin{figure}[!ht]
\begin{center}
\includegraphics[width=0.85\textwidth]{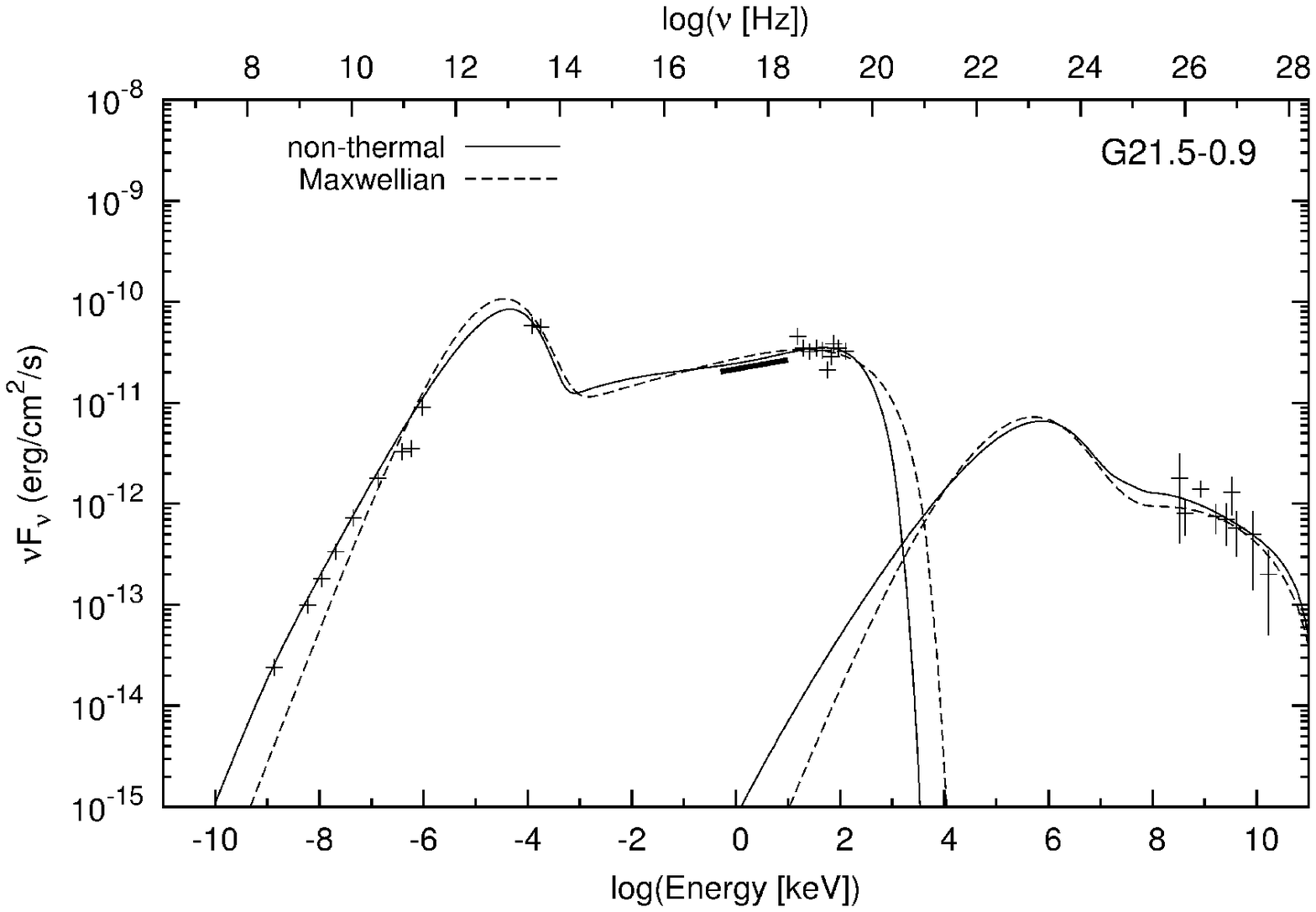}
\caption{Model prediction for the young nebula G21.5--0.9.  Radio data are taken from \citet{Goss1970}, \citet{Becker1975}, \citet{Morsi1987}, \citet{Salter1989b, Salter1989a}, \citet{Bock2001}, \citet{Bandiera2001}, \citet{Bietenholz2008}, and \citet{Bietenholz2011}, infra-red data from \citet{Gallant1999}, X-ray data from \citet{Slane2000}, \citet{DeRosa2009}, and the INTEGRAL Science Data Centre (http://www.isdc.unige.ch/heavens webapp/integral), and TeV data from \citet{Dejager2008b}.  The IC prediction is calculated using both the CMBR and an SSC component, although the former component leads to a negligible flux.}
\label{fig:G21.5}
\end{center}
\end{figure}

With a spin-down luminosity of $L=3.3 \times 10^{37}\,\text{erg}\,\text{s}^{-1}$ \citep{Camilo2006}, the pulsar in the SNR G21.5--0.9 is one of the most energetic pulsars in the Galaxy.  The PWN is located at a distance of $4.8\,\text{kpc}$ \citep{Tian2008}, with an estimated age of $870\,\text{yr}$ \citep{Bietenholz2008}.  Radio \citep[e.g.,][]{Goss1970, Becker1975, Morsi1987, Salter1989a, Salter1989b, Bock2001, Bandiera2001, Bietenholz2008, Bietenholz2011}, IR \citep[e.g.,][]{Gallant1999, Zajczyk2012}, and X-ray \citep{Slane2000, DeRosa2009, Tsujimoto2011} observations of the PWN show a bright nebula with a radius of $\sim 40''$.  The nebula is embedded in diffuse X-ray emission, believed to be the result of dust-scattered X-rays from the PWN \citep{Bocchino2005}.  At GeV energies, \emph{Fermi}-LAT only detected upper limits for the nebula \citep{Ackermann2011}, while a detection at TeV energies has been reported by the H.E.S.S. Collaboration \citep{Djannati2007, Dejager2008b}.

For the modelling of G21.5--0.9, a number of constraints on the parameters follow from observations.  As the age and present-day luminosity are known, the initial luminosity is fixed for a choice of $\tau$ using Equation (\ref{eq:L_t}).  For the spin-down time scale, the value of $\tau=3\,\text{kyr}$ estimated by \citet{Dejager2009b} is used.  The derived distance to the source implies a PWN radius of $R\s{pwn}=0.93\,\text{pc}$.  The PWN is too young to have interacted with the reverse shock, and must therefore still be in the first expansion phase.  With the age and size of the PWN taken into account, the expansion rate in Equation (\ref{eq:R_t}) has the value $r_1=1.02$, leading to a present-day expansion velocity of $v\s{age}=1060\,\text{km}\,\text{s}^{-1}$.      

Figure \ref{fig:G21.5} shows the model prediction for the non-thermal radiation spectra, with the derived parameters listed in Table \ref{tab:parameters}.  In order to find an agreement between the TeV data and the model prediction, it is necessary to include synchrotron self-Compton scattering in the model.  \citet{Atoyan1996} calculated that the density of the synchrotron photon field is
\bb\label{eq:ssc}
n_{\rm{ssc}} = \frac{Q_{\rm{syn}}}{4\pi R_{\rm{syn}}c}\frac{\bar{U}}{h\nu},
\ee       
where $Q\s{syn}$ is the synchrotron emissivity, $R_{\rm{syn}}$ the radius within which most of the synchrotron emission is produced, and $\bar{U} \simeq 2.24$.  Additionally, photons from the CMBR  were also taken into account, but it was found that this radiation field is significantly less important than the SSC component. 

From the model a present-day average magnetic field of $B\s{age}=230\,\mu\text{G}$ is derived.  This is comparable to the value of $B\s{age}=300\,\mu\text{G}$ inferred for the $\sim 1\,\text{kyr}$ old Crab Nebula \citep[e.g.,][]{Trimble1982}.  To obtain the model prediction presented in Figure \ref{fig:G21.5}, the diffusion coefficient should not be larger than $\kappa= 2.2 \times 10^{25} E\s{TeV}\,\text{cm}^2\,\text{s}^{-1}$.  For Vela X with an average magnetic field of $B\s{age}=5\,\mu\text{G}$ \citep[e.g.,][]{Dejager2008c}, a value of $\kappa=10^{27} E\s{TeV}\,\text{cm}^2\,\text{s}^{-1}$ has been estimated by \cite{Hinton2011} to explain the absence of particles with an energy $E_e > 100$ GeV from the relic PWN observed by \emph{Fermi}-LAT.  Treating the magnetic field and diffusion coefficient of Vela X as fiducial values, the scaling $\kappa \propto 1/B$ implies that the magnetic field $B\s{age}=230\,\mu\text{G}$ should lead to $\kappa= 2.2 \times 10^{25} E\s{TeV}\,\text{cm}^2\,\text{s}^{-1}$, in agreement with the value derived from the model.  In terms of the Bohm diffusion coefficient (\ref{eq:kappa_bohm}), the derived diffusion coefficient has the value of $\kappa=390\kappa\s{Bohm}$.    

Among the parameters derived for the nebula G21.5--0.9 is the ratio of conversion efficiencies $\eta\s{R}/\eta\s{X}=5.4$, significantly smaller than the ratio $\eta\s{R}/\eta\s{X}=116-150$ derived by \citet{Dejager2008c} for Vela X.  The ratio of magnetic to particle energy is found to be $\sigma=0.18$, larger than the value of $\sigma \sim 0.003$ calculated by \citet{Kennel1984b} for the Crab Nebula.  The well-known steady-state magnetohydrodynamic model of \cite{Kennel1984a} predicts a radial velocity that is almost independent of $r$ when $\sigma=0.25$.  Based on this result, Equation (\ref{eq:E_dot_ad2}) should be a reasonable approximation for the adiabatic losses when $\sigma=0.18$, as the velocity in the largest part of the PWN will not differ significantly from the expansion velocity.

To illustrate the effect of a continuous/discontinuous source spectrum on the derived parameters, one need only consider the modeling results of \citet{Tanaka2010}.  Using a continuous source spectrum and a model similar to the present one, these authors derived a smaller present-day magnetic field of $B\s{age}\le 64\,\mu\text{G}$ for G21.5--0.9.  However, \citet{Tanaka2010} were unable to predict the $1-10$ keV X-ray observations, while the present model with the different normalisation constants is a very good description of the broadband spectra.  Furthermore, it was found that a magnetic field much smaller than $B\s{age} = 230\,\mu\text{G}$ cannot be used in the present model, as this leads to the requirement $\eta\s{R} > 1$.  

Following the discussion presented in Section \ref{sec:two_com}, the non-thermal emission from G21.5--0.9 is also modelled using a Maxwellian spectrum with a power-law tail
\bb\label{eq:thermal}
Q(E_e,t) = \begin{cases}
Q_{\rm{T}}\left(E_e/E_{\rm{ts}}\right)\exp\left[-E_e/E_{\rm{ts}}\right], & \text{for all } E_e\\
Q_{\rm{N}}\exp\left[(-E-E_{\max})/\Delta E_{\max}\right]\left(E_e/E_{\rm{ts}}\right)^{-\alpha_N}, & \text{if } E_{\rm{b}}<E_e\le E_{\max} 
\end{cases},
\ee
where $Q\s{T}$ and $Q\s{N}$ respectively represent the normalisation constants for the thermal and non-thermal components, and $E_{\rm{ts}}=0.26(\gamma/10^6)\,\text{TeV}$.  Here $\gamma$ represents the Lorentz factor of the electrons upstream of the termination shock, while $E_{\rm{b}}=7E_{\rm{ts}}$.  Note that Equation (\ref{eq:thermal}) is a slightly modified \citet{Spitkovsky2008} spectrum introduced by \citet{Fang2010}.  The aim is not to model G21.5--0.9 time-dependently using Equation (\ref{eq:thermal}), but merely to determine whether a Maxwellian spectrum can be used to explain the radio/GeV data. 

Figure \ref{fig:G21.5} shows that Equation (\ref{eq:thermal}) can also be used to model the broadband data, except at radio frequencies where a much harder spectrum is predicted.  From this modeling values for $E\s{min}$ and $E\s{max}$ are derived that are similar to the values listed in Table \ref{tab:parameters}.  Other parameters that are derived include a Lorentz factor of $\gamma=5\times 10^4$, implying a break energy of $E\s{b}=0.09\,\text{TeV}$, along with $\alpha_N=2.7$ and $\Delta E_{\max}=160\,\text{TeV}$ for the power-law tail.  These last two values are comparable to the values of $\alpha_N=2.5$ and $\Delta E_{\max}=100\,\text{TeV}$ predicted by \cite{Spitkovsky2008}.


\subsection{HESS J1427--608}

\begin{figure}[!ht]
\begin{center}
\includegraphics[width=0.85\textwidth]{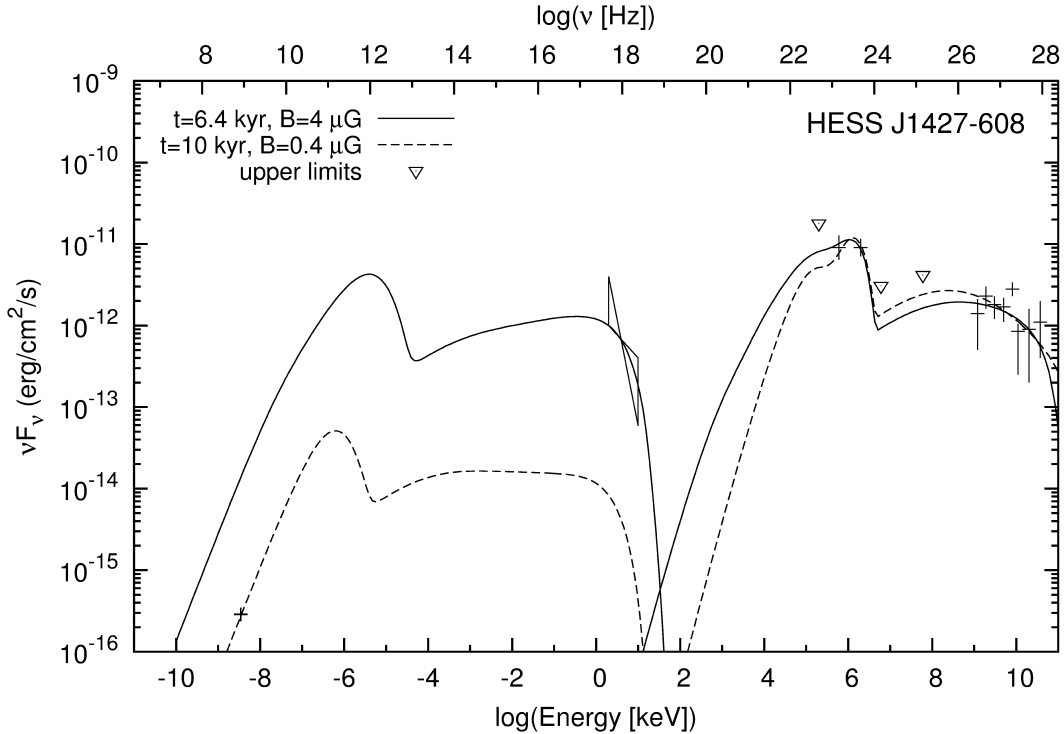}
\caption{Model prediction for the unidentified source HESS J1427--608.  The radio data is taken from \cite{Murphy2007}, the X-ray data from \cite{Fujinaga2012}, the \emph{Fermi} data from \cite{Nolan2012}, and the TeV data from \cite{H_Aharonian2008}.  For this source the GeV/TeV spectrum is produced by the IC scattering of both the CMBR and an IR component.}
\label{fig:1427}
\end{center}
\end{figure}

One of the sources discovered in a H.E.S.S. Galactic Plane Survey is HESS J1427--608, with an intrinsic source extension of $2'.4-4'.8$ \citep{H_Aharonian2008}.  Observations by \emph{Fermi}-LAT detected an associated GeV source, 2FGL J1427.6-6048 \citep{Nolan2012}, while  \cite{Fujinaga2012} recently reported an X-ray detection with \emph{Suzaku} in the $2-10$ keV band.  The X-ray emission is spatially coincident with the TeV emission, and has a radius of $2'$.  Based on the well-known fact that the VHE nebula is typically larger than the X-ray nebula \citep[e.g.,][]{Kargaltsev2010}, it was suggested by \cite{Fujinaga2012} that the \emph{Suzaku} observation represents the X-ray counterpart to HESS J1427--608.  Using the X-ray observations, \cite{Fujinaga2012} estimated that the source is located at a distance of $d\sim 11\,\text{kpc}$, and has an age of $t\s{pwn} \sim 6.4\,\text{kyr}$.  A possible radio counterpart, MGPS J142755-605038 with a radius of $0.56'-0.77'$, has been observed in a Molonglo Sky Survey \citep{Murphy2007}.  As the radio nebula of a PWN is typically larger than the X-ray nebula \citep[e.g.,][]{Gaensler2006}, it is difficult to simultaneously associate both the radio and \emph{Suzaku} sources with the VHE emission, assuming that HESS J1427--608 is indeed a PWN, and that the \emph{Suzaku} detection represents the X-ray nebula.  This incompatibility is also strongly underlined by the model.

For the modeling a spherical source with a radius of $2'.4$ is used, along with the distance and age estimates derived by \cite{Fujinaga2012}.  The observed size and estimated distance lead to a radius of $R\s{pwn}=7.7\,\text{pc}$.  For a $\sim 6\,\text{kyr}$ source it is entirely possible that the PWN has not yet interacted with the reverse shock, and it is thus assumed that HESS J1427--608 is still in the first expansion phase.  The rate of expansion is $r_1=1.03$, leading to present-day expansion velocity of $v\s{age}=1200\,\text{km}\,\text{s}^{-1}$.  To model the GeV/TeV data it is necessary to not only include IC scattering of the CMBR, but also scattering of an IR photon field.  The energy spectrum of the IR photons is taken as a black-body spectrum with a temperature of $T=50\,\text{K}$ and an energy density of $U\s{IC}=2\,\text{eV}\,\text{cm}^{-3}$.  This is comparable to the values of $T=46\,\text{K}$ and $U\s{IC}=5\,\text{eV}\,\text{cm}^{-3}$ used by \cite{Zhang2008} to model the TeV data of MSH 15-52 and HESS J1825-137.  

Apart from the radio measurement, Figure \ref{fig:1427} shows that the model prediction is in good agreement with the data if adiabatic losses are neglected.  The scenario presented in Figure \ref{fig:1427} requires a large initial luminosity ($L_0=1.2\times 10^{39}\,\text{erg}\,\text{s}^{-1}$) and spin-down time scale ($\tau=3\,\text{kyr}$), leading to a present-day luminosity of $L=1.2 \times 10^{38}\,\text{erg}\,\text{s}^{-1}$.  From the model prediction a present-day magnetic field of $B\s{age}=4.2\,\mu\text{G}$ is derived.  The ratio of particle conversion efficiencies is $\eta\s{R}/\eta\s{X}=13.8$, comparable to the value of $\eta\s{R}/\eta\s{X}=5.4$ derived for G21.5--0.9.  Moreover, the model predicts that the ratio of magnetic to particle energy is $\sigma = 4\times 10^{-4}$.  The small $\sigma$ value derived implies that the energy content in this source is predominantly stored in the particles, in contrast to G21.5--0.9 ($\sigma=0.18$) where the electromagnetic energy is an important fraction of the total energy.  For the escape losses, the model predicts a present-day diffusion coefficient of $\kappa=10^{26} E\s{TeV}\,\text{cm}^2\,\text{s}^{-1}$, or equivalently, $\kappa=30\kappa\s{Bohm}$.    

\citet{Vorster2013b} calculated that when $\sigma<0.01$, the radial convection velocity in the PWN decreases as $v\propto 1/r^2$.  It follows from Equation (\ref{eq:ad_r}) that the particles will not be subjected to adiabatic losses for such a profile, thereby motivating the neglect of this energy loss process from the modeling.  An agreement between the model and data could also be found with adiabatic losses included.  In this scenario a marginally larger present-day magnetic field of $B\s{age}=3.9\,\mu\text{G}$ is derived.  The largest difference, compared to the scenario presented in Figure \ref{fig:1427}, is that a very large initial luminosity of $L_0=6.5\times 10^{39}$ is required to make up for the adiabatic losses suffered by the low-energy particles.      

The present-day value of $L=1.2 \times 10^{38}\,\text{erg}\,\text{s}^{-1}$ predicted by the model (with adiabatic losses neglected) is remarkably similar to a pulsar that has recently been detected by \cite{Arzoumanian2011} near the Galactic plane.  The authors measured a value of $L=1.2\times 10^{38}\,\text{erg}\,\text{s}^{-1}$ for PSR J2022+3842, and estimated the age of the source to be $t\s{age}=8.9\,\text{kyr}$.  Although there is some uncertainty regarding the distance to PSR J2022+3842, \cite{Arzoumanian2011} placed the pulsar at $d=10\,\text{kpc}$.  Furthermore, the authors also detected a very faint elliptical PWN in X-rays with the total dimensions of $29 \times 35\,\text{pc}$.  The only difference between this PWN and HESS J1427--608 is that a bright radio nebula, G76.9+1.0 \citep{Landecker1993}, is associated with the X-ray PWN.  Although the nature of G76.9+1.0 is not entirely clear, \cite{Landecker1993} argued that the filled centre of the radio source is more indicative of a PWN than an SNR.  However, the authors derived a spectral index of $\alpha_R=0.62$, much steeper than the values $\alpha_R=0-0.3$ typically associated with PWNe \citep[e.g.,][]{Weiler1980}.

Even though the values derived from the model prediction shown in Figure \ref{fig:1427} are compatible with PWN parameters, the over-prediction of the radio data makes it difficult to unambiguously accept this scenario.  The model predicts a bright radio source that has thus far not been observed in the region of the sky spatially coincident with the position of HESS J1527-608.  A model prediction compatible with the radio data can be obtained from the scenario presented above, provided that the minimum electron energy is $E\s{min} > 0.1\,\text{TeV}$.  This seems an unnatural high value, and this solution is therefore disfavoured.  

An alternative solution would be to decrease the value of the magnetic field.  Figure \ref{fig:1427} also shows a scenario where an agreement between the model prediction and radio data has been obtained using a present-day magnetic field of $B\s{age}=0.42\,\mu\text{G}$, with the parameters derived from this scenario listed in Table \ref{tab:parameters}.  In order for the magnetic field to reach such a low value, the PWN must be older than the value of $t\s{age}=6.4\,\text{kyr}$ estimated by \cite{Fujinaga2012}, and the larger value of $t\s{age}=10\,\text{kyr}$ is chosen as the age of the PWN.

For this alternative scenario, a smaller (compared to the $B\s{age}=4.2\,\mu\text{G}$ scenario) expansion rate of $r_1=0.96$ is derived, leading to a present-day expansion velocity of $v\s{age}=720\,\text{km}\,\text{s}^{-1}$.  This scenario requires a smaller initial luminosity ($L_0=5.5\times 10^{38}\,\text{erg}\,\text{s}^{-1}$), leading to a present-day luminosity of $L=2.9 \times 10^{37}\,\text{erg}\,\text{s}^{-1}$.  The ratio of the conversion efficiencies is $\eta\s{R}/\eta\s{X}=4.5$, while the ratio of particle to magnetic energy is $\sigma =10^{-5}$.  Lastly, a diffusion coefficient of $\kappa= 5.6 \times 10^{25} E\s{TeV}\,\text{cm}^2\,\text{s}^{-1}$ is predicted by the model, or $\kappa=2.3\kappa\s{Bohm}$.  Note that the prediction of the radio data requires a large minimum energy ($E\s{min}=0.1\,\text{TeV}$) that can be reduced to $E\s{min}=10^{-2}\,\text{TeV}$ if the magnetic field is decreased to $B\s{age}=0.1\,\mu\text{G}$.  While these parameters may lead to an acceptable agreement between the model and radio data, Figure \ref{fig:1427} shows that this scenario significantly under-predicts the \emph{Suzaku} spectrum.  This discrepancy can be explained if the X-ray observations are not related to the TeV nebula.   

To understand why the model has difficulty in predicting both the radio and X-ray synchrotron data, one need only consider the two quantities that eventually determine the synchrotron flux: the number of particles that produce the non-thermal emission, and the magnetic field strength.  As the magnetic field is the same for both the radio and X-ray producing particles, the only way to predict both the radio and X-ray data would be to increase the number of high-energy particles.  However, it follows from Figure \ref{fig:test} that this would also increase the TeV flux, leading to an over-prediction of the H.E.S.S. data.


\subsection{HESS J1507--622}

\begin{figure}[!t]
\begin{center}
\includegraphics[width=0.85\textwidth]{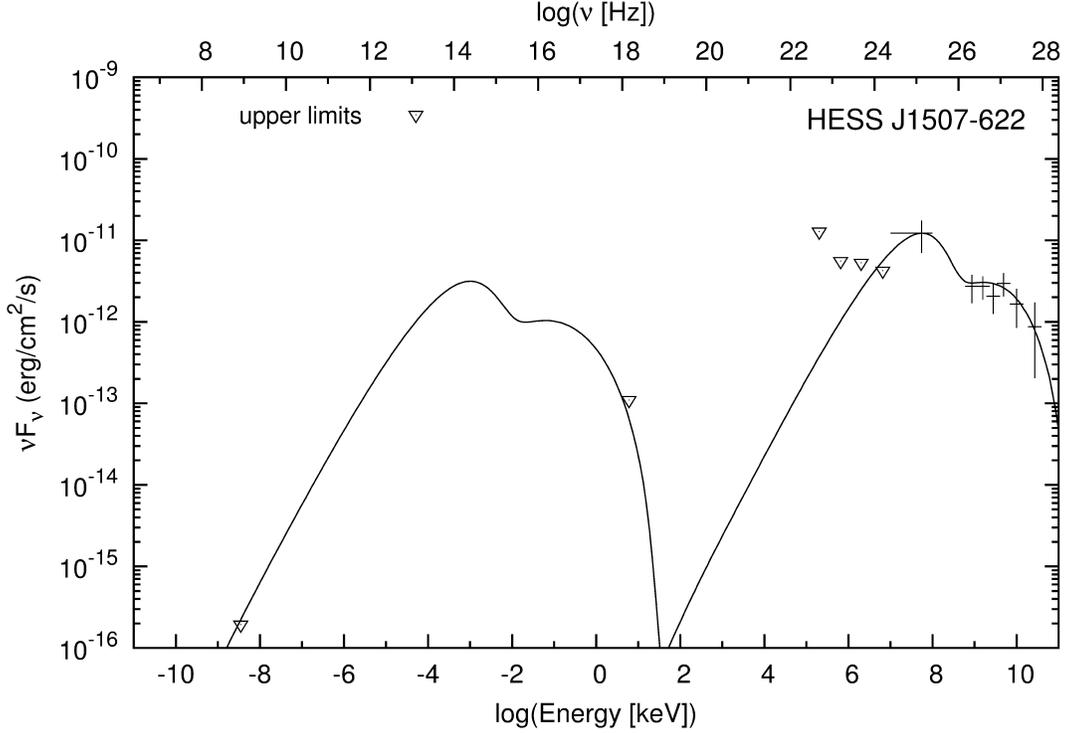}
\caption{Model prediction for the unidentified source HESS J1507--622.  The radio upper limit is taken from \cite{Green1999}, and the GeV data from \cite{Nolan2012}.  The X-ray upper limit and the TeV data are taken from the \cite{H_Aharonian2011}.  The IC spectrum is produced by only taking into account the scattering of CMBR photons.}
\label{fig:1507}
\end{center}
\end{figure}

Also discovered in a H.E.S.S. Galactic Plane Survey is the bright ($\sim 8\%$ of the Crab flux) VHE source HESS J1507--622, with a radius of $\sim 9'$ \citep{H_Aharonian2011}.  A possible synchrotron counterpart may be provided by the extended, diffuse X-ray emission (with a radius of $10''-13''$) observed with \emph{Chandra} \citep{H_Aharonian2011}.  However, the identification of this X-ray source with the VHE emission region remains inconclusive \citep{H_Aharonian2011}.  An additional synchrotron upper limit is provided by the source MGPS J150850-621025 discovered in a Molonglo Galactic Plane Survey \citep{Green1999}.  Lastly, a GeV counterpart, 2FGL J1507.0-6223, has recently been discovered by the \emph{Fermi}-LAT Collaboration and is reported in \cite{Nolan2012}. 

HESS J1507--622 is unique in the sense that it lies $\sim 3^{\circ}.5$ from the Galactic plane, whereas all other unidentified source lie within $\pm 1^{\circ}$ from the Galactic equator.  Most Galactic VHE sources are connected to young stellar populations (located in the disk), and one would therefore not expect a bright VHE source at the observed position.  Furthermore, the absence of a bright X-ray counterpart is surprising as the comparably low hydrogen column density at $\sim 3.5^{\circ}$ leads to a considerably lower absorption of X-rays, as well as reduced background emission \citep{H_Aharonian2011}.  

To explain the uniqueness of the source, the \cite{H_Aharonian2011} considered a number of possible scenarios.  On the one hand, the absence of counterparts, especially in X-rays, suggests a hadronic scenario.  Given the low density of target material off the Galactic plane \citep[see e.g.,][]{Lockman1984}, this scenario was disfavoured by the \cite{H_Aharonian2011} unless the source could be placed at a very small distance of $d < 1\,\text{kpc}$.  Although unlikely, the hadronic scenario can not be fully excluded \citep{H_Aharonian2011}.  An alternative scenario is that HESS J1507--622 is an ancient PWN.  As a result of its small angular extension, the leptonic scenario would place the source at a distance of $d > 6\,\text{kpc}$ \citep{H_Aharonian2011}.  

For the modeling the hints of diffuse X-ray emission detected by \emph{Chandra} are taken as an upper limit.  The source is placed at a distance of $d=6\,\text{kpc}$, leading to a radius of $R\s{pwn}=15.7\,\text{pc}$.  For the first modelling attempts, it was assumed that HESS J1507--622 is still in the initial expansion phase.  In order to model the GeV data, this scenario requires a break energy of $E\s{b}=5\,\text{TeV}$.  This is an order of magnitude larger than the values derived for G21.5--0.9 ($E\s{b}=0.1\,\text{TeV}$) and HESS J1427--608 ($E\s{b}=0.18\,\text{TeV}$).  Values similar to those presented in Table \ref{tab:parameters} have also been derived for a number of known PWNe, including the $\sim 21\,\text{kyr}$ old nebula HESS J1825-137.  \cite{Zhang2008} found that $E\s{b}\le 0.15\,\text{TeV}$, while \cite{Tanaka2011} derived values that where $E\s{b}\le 0.3\,\text{TeV}$.  The exception to the results of \cite{Tanaka2011} is Kes 75, where the authors derived a possible value of $E\s{b}=2.6\,\text{TeV}$.  Using a Maxwellian source spectrum, \cite{Fang2010} derived the even smaller values of $E\s{b}=0.02-0.09\,\text{TeV}$. 

Although it is not excluded that such a large break energy is the result of shock acceleration, an alternative scenario is favoured in the present paper where HESS J1507--622 has been compressed by the reverse shock.  As $\dot{E_e}/E_e$ is constant in Equation (\ref{eq:E_dot_ad2}), the effect of adiabatic losses is to shift the electron spectrum to lower energies without affecting the spectral shape, as illustrated in Figure \ref{fig:test}.  During the compression phase, the exact opposite will occur.  The particles will be subjected to adiabatic heating, causing the electron spectrum to shift to higher energies, thereby leading to an increase in the value of $E\s{b}$.    

At an offset of $3^{\circ}.5$ one would expect the ISM to have a lower density compared to the Galactic plane, and from the reverse shock time scale given by Equation (\ref{eq:t_rev}), it follows that smaller ISM densities lead to larger time scales.  Inserting the values of $E_{\rm{ej}}=10^{51}$ erg, $M_{\rm{ej}}=9 M_{\odot}$, and $n_{\rm{ism}}= 0.1 \mbox{ cm}^{-3}$ into Equation (\ref{eq:t_rev}) leads to an estimate of $t_{\rm{rs}} = 19.6\,\text{kyr}$.  For the compression scenario the value of $t_{\rm{rs}} = 20\,\text{kyr}$ is used, while the compression phase is chosen to last for $4\,\text{kyr}$.  It is assumed that the nebula has not yet entered the second expansion phase, implying that the current age of the PWN is $t\s{age}=24\,\text{kyr}$.  In the initial phase the PWN expands to a radius of $R\s{pwn}=20\,\text{pc}$ with a constant velocity of $v\s{pwn}=980\,\text{km}\,\text{s}^{-1}$.  In the next phase the interaction with the reverse shock compresses the PWN, thereby causing $v\s{pwn}$ to reverse direction.  During this compression phase $R\s{pwn}$ initially moves inward with a velocity of $v\s{pwn}=1300\,\text{km}\,\text{s}^{-1}$, reducing to $v\s{pwn}=850\,\text{km}\,\text{s}^{-1}$ after $t=24\,\text{kyr}$.  Given the offset from the Galactic plane, it seems reasonable to assume that the ISM is homogeneous.  This will lead to a symmetric reverse shock and a preservation of the spherical nature of the PWN.  Furthermore, one would not expect any photon field other than the CMBR to be present at the position of HESS J1507--622, and only this component is taken into account.  

The model prediction resulting from the compression scenario is shown in Figure \ref{fig:1507}, with the derived parameters listed in Table \ref{tab:parameters}.  With the compression taken into account, the break energy is reduced to $E\s{b}=0.5\,\text{TeV}$.  Note that this is the break energy of the source spectrum, while the particle spectrum in the PWN has a break at $5\,\text{TeV}$.  The effect of the compression is also reflected in the derived value of $E\s{min}$.  Comparing Figures \ref{fig:1427} and \ref{fig:1507} shows that the radio data for the two unidentified TeV sources are very similar, yet the values of $E\s{min}$ listed in Table \ref{tab:parameters} differ significantly between the two sources.  Although the model predicts a value of $E\s{min}=10^{-3}\,\text{TeV}$ at the termination shock of HESS J1507--622, the compression also increases the minimum electron energy in the nebula, thereby making it possible for the radio synchrotron spectrum to be compatible with the upper limit.     

Other parameters derived include a relatively large initial luminosity ($L_0=1.2\times 10^{38}\,\text{erg}\,\text{s}^{-1}$) and a short spin-down time scale ($\tau=0.5\,\text{kyr}$).  As the expansion of the PWN is driven by the continual injection of the pulsar's spin-down energy into the nebula, the small $\tau$ value (compared to the other two scenarios investigated) implies that the energy input declines rapidly with time, thereby motivating the constant expansion velocity in the initial phase.  A present-day magnetic field of $B\s{age}=1.7\,\mu\text{G}$ is derived from the model, larger than the value of $B\s{age}=0.5\,\mu\text{G}$ estimated by \cite{Tibolla2011} and the \cite{H_Aharonian2011}.  Additionally, the ratios $\eta\s{R}/\eta\s{X}=4.7$ and $\sigma=0.03$ are derived.  A present-day diffusion coefficient of $\kappa=1.5 \times 10^{26} E\s{TeV}\,\text{cm}^2\,\text{s}^{-1}$ is derived, comparable to the value derived for HESS J1427--608.  In terms of the Bohm coefficient (\ref{eq:kappa_bohm}), the derived diffusion coefficient has the value of $\kappa=20\kappa\s{Bohm}$.  For the expansion phase, adiabatic losses are calculated using Equation (\ref{eq:E_dot_ad2}), while the compression phase requires adiabatic heating that is ten times larger than that predicted by Equation (\ref{eq:E_dot_ad2}).


\section{Discussion and Conclusions}

In this paper a time-dependent PWN particle evolution model is presented and applied to the young PWN G21.5--0.9, as well as to the unidentified TeV sources HESS J1427--608 and HESS J1507--622.  For the three sources sets of parameters are derived that are reasonable within a PWN framework, thereby strengthening the argument that HESS J1427--608 and HESS J1507--622 can be identified as PWNe.  The robustness of the derived parameter sets was tested by considering a large number of alternative scenarios.  It was found that markedly changing the values of the parameters, compared to those given in Table \ref{tab:parameters}, leads to model predictions that are not compatible with the observations.  

As discussed in \citet{Possenti2002}, observations indicate that the X-ray luminosity of the PWN decreases with age, while \citet{Mattana2009} found that there is no correlation between the pulsar's spin-down luminosity and the observed TeV luminosity of the PWN.  The corollary is that the PWN will remain bright at TeV energies, even if the pulsar's spin-down luminosity has reached an undetectable level \citep[e.g.,][]{Dejager2009b,H_Aharonian2011}.   This evolutionary trend is also predicted by the model, as the synchrotron luminosity of the evolved PWNe fades away below the sensitivity of the current generation of X-ray satellites, while nevertheless remaining bright at TeV energies.  

Apart from detecting pulsars that can be associated with the unidentified sources, additional multi-wavelength observations may further strengthen the PWN identification.  A characteristic of PWNe is that electrons and positrons are responsible for the observed non-thermal emission.  Although it will not directly confirm an unidentified TeV source as a PWN, the detection of a $511\,\text{keV}$ annihilation line by future sub-MeV experiments, e.g., the proposed GRIPS satellite \citep{Greiner2012}, will at least indicate that the particles responsible for the TeV emission in the unidentified sources are leptonic \citep{Tibolla2011b}.  

As such, only a few alternative explanations for the unidentified TeV sources have thus far been proposed, including the suggestion by \citet{Yamazaki2006} that these sources can be associated with old SNRs.  Arguing that SNRs can only confine multi-TeV particles for a very short period ($t \lesssim 1\,\text{kyr}$), the aforementioned proposal has however been questioned by \citet{Gabici2007}.  As a second alternative, \citet{Gabici2007} have suggested that the unidentified sources can still be identified with SNRs if multi-TeV particles that have escaped from the remnant interact with nearby dense clouds.  This scenario seems unlikely for the unidentified sources discovered so far, given the absence of dense molecular clouds spatially coincident with most of these sources. Given its location above the Galactic plane, this is particularly true for HESS J1507--622. 

Motivated by observations, a broken power-law is used as the source spectrum for the electrons injected into the PWN at the termination shock.  In contrast to previous PWN models of a similar nature \citep[see e.g.,][]{Zhang2008,Tanaka2010}, the source spectrum in the present model has a discontinuity in intensity at the transition between the low and high-energy components.  The choice of a discontinuous source spectrum leads to a better model prediction of the data at all wavelengths, in contrast to a continuous one.  A similar conclusion has also been drawn by \cite{Dejager2008c} from their modeling of Vela X.  As a discontinuous spectrum is also required for the young ($t\s{age}\sim 1\,\text{kyr}$) nebula G21.5--0.9, the discrepancy between the two components cannot be an artifact of PWN evolution.  A characteristic of the discontinuous spectrum is that a particle conversion efficiency must be specified for both the low ($\eta\s{R}$) and high-energy ($\eta\s{X}$) components, with a ratio of $\eta\s{R}/\eta\s{X}\sim 4.5-5.4$ derived for the three sources.  

The data for G21.5--0.9 were also modeled using a Maxwellian source spectrum with a non-thermal tail.  The aim was not to model the evolution of the PWN time-dependently, but rather to illustrate that a Maxwellian source spectrum can be used to predict the data.  Although the Maxwellian spectrum supplies a natural explanation for the ratio $\eta\s{R} /\eta\s{X} > 1$, a synchrotron radio spectrum is predicted that is harder than the one observed.  However, as discussed in Section \ref{sec:two_com}, the results of \citet{Sironi2011} indicate that magnetic reconnection at the termination shock can accelerate particles, leading to a modification of the Maxwellian that would produce a softer radio synchrotron spectrum.  

For HESS J1427--608 two possible scenarios were investigated.  In the first, a present-day magnetic field of $B\s{age}=4.2\,\mu\text{G}$ is derived, along with a present-day luminosity of $L=1.2\times 10^{38}\,\text{erg}\,\text{s}^{-1}$.  This scenario predicts PWN values that are very similar to the recently discovered pulsar PSR J2022+3842 and its associated PWN \citep{Arzoumanian2011}.  However, the $B\s{age}=4.2\,\mu\text{G}$ scenario predicts a bright radio nebula that has thus far not been observed.  An alternative scenario is considered where the magnetic field in the nebula has evolved to a very low value of $B\s{age}=0.4\,\mu\text{G}$, leading to a synchrotron spectrum that is compatible with radio upper limits.  However, this scenario significantly under-predicts the \emph{Suzaku} observations presented by \citet{Fujinaga2012}.  The $B\s{age}=0.4\,\mu\text{G}$ scenario represents an ancient PWN as the very small present-day magnetic field leads to a low level of synchrotron emission, while still remaining bright at GeV/TeV energies.  The fact that the model cannot simultaneously predict both the radio and X-ray observations strengthens the idea that one of the two synchrotron sources is not a plausible counterpart to HESS J1427--608.

Assuming that HESS J1507--622 is still in the initial expansion phase, the energy value where the electron spectrum transitions from the low to high-energy components was found to be $E\s{b}=5\,\text{TeV}$.  This is an order of magnitude higher than the values derived for G21.5--0.9, HESS J1427--608, and a number of other PWNe \citep[e.g.,][]{Zhang2008, Fang2010, Tanaka2011}.  Therefore, in this paper a scenario is favored where HESS J1507--622 has been compressed by the reverse shock of the SNR.  As a result of the adiabatic heating, the break energy has been increased from $E\s{b}=0.5\,\text{TeV}$ to $E\s{b}=5\,\text{TeV}$ in the nebula.  This compression scenario leads to a derived age and present-day magnetic field of $t\s{age}=24\,\text{kyr}$ and $B\s{age}=1.7\,\mu\text{G}$ respectively .  If the faint X-ray source is indeed the synchrotron counterpart to the VHE nebula, the present modeling results strengthens the argument for identifying HESS J1507--622 as a PWN.


It is tempting to use the derived parameters listed in Table \ref{tab:parameters} as a guideline of PWN evolution.  It is reiterated that the model does not include a time-dependence in a number of parameters (e.g., $\eta\s{R}$, $\eta\s{X}$, $E\s{b}$), and no comment can be made regarding the temporal evolution of these parameters.  However, the model does allow for the derivation of additional parameters, most notably the ratio $\eta\s{R} /\eta\s{X}$, that may be unique to PWNe.  This requires that the current model be applied to a large number of known PWNe allowing for the derivation of a statistically significant set of parameters.  A follow-up paper addressing this issue is currently planned.

\begin{acknowledgements}
The authors acknowledge and honor the memory of a great scientist, a wonderful person, and the one who introduced us to the idea of ancient pulsar wind nebulae: O.C. de Jager. Okkie, working with you was a great honor and we will miss you very much.  The authors thank the anonymous referee for his/her useful comments, as well as Karl Mannheim for his helpful discussions.
\end{acknowledgements}

\bibliographystyle{aa}
\bibliography{References_Vorster_Tibolla}

\end{document}